\documentstyle[eqsecnum,rotating,epsfig,aps]{revtex}
 \begin{document}
\newcommand{\str}{\rule{0ex}{2.7ex}}  
%
\newcommand{\pr}{Phys.\ Rev.\ }
\newcommand{\prp}{Phys.\ Rep.\  }
\newcommand{\np}{Nucl.\ Phys.\ B }
\newcommand{\zp}{Z.\ Phys.\ C }
\newcommand{\beq}{\begin{equation}}
\newcommand{\eeq}{\end{equation}}
\newcommand{\bea}{\begin{eqnarray}}
\newcommand{\eea}{\end{eqnarray}}
%
%
\newbox\hdbox%
\newcount\hdrows%
\newcount\multispancount%
\newcount\ncase%
\newcount\ncols
\newcount\nrows%
\newcount\nspan%
\newcount\ntemp%
\newdimen\hdsize%
\newdimen\newhdsize%
\newdimen\parasize%
\newdimen\spreadwidth%
\newdimen\thicksize%
\newdimen\thinsize%
\newdimen\tablewidth%
\newif\ifcentertables%
\newif\ifendsize%
\newif\iffirstrow%
\newif\iftableinfo%
\newtoks\dbt%
\newtoks\hdtks%
\newtoks\savetks%
\newtoks\tableLETtokens%
\newtoks\tabletokens%
\newtoks\widthspec%
%
%
\immediate\write15{%
CP SMSG GJMSINK TEXTABLE --> TABLE MACROS V. 851121 JOB = \jobname%
}%
%
%
\tableinfotrue%
\catcode`\@=11
\def\out#1{\immediate\write16{#1}}
%
%
\def\tstrut{\vrule height3.1ex depth1.2ex width0pt}%
\def\and{\char`\&}
\def\tablerule{\noalign{\hrule height\thinsize depth0pt}}%
\thicksize=1.5pt
\thinsize=0.6pt
\def\thickrule{\noalign{\hrule height\thicksize depth0pt}}%
\def\hrulefill{\leaders\hrule\hfill}%
\def\bigrulefill{\leaders\hrule height\thicksize depth0pt \hfill}%
\def\ctr#1{\hfil\ #1\hfil}%
\def\altctr#1{\hfil #1\hfil}%
\def\vctr#1{\hfil\vbox to0pt{\vss\hbox{#1}\vss}\hfil}%
%
%
\tablewidth=-\maxdimen%
\spreadwidth=-\maxdimen%
\def\tabskipglue{0pt plus 1fil minus 1fil}%
%
%
\centertablestrue%
\def\centeredtables{%
   \centertablestrue%
}%
\def\noncenteredtables{%
   \centertablesfalse%
}%
%
%
\parasize=4in%
\long\def\para#1{
   {%
      \vtop{%
         \hsize=\parasize%
         \baselineskip14pt%
         \lineskip1pt%
         \lineskiplimit1pt%
         \noindent #1%
         \vrule width0pt depth6pt%
      }%
   }%
}%
\gdef\ARGS{########}
\gdef\headerARGS{####}
\def\@mpersand{&}
{\catcode`\|=13
\gdef\letbarzero{\let|0}
\gdef\letbartab{\def|{&&}}%
\gdef\letvbbar{\let\vb|}%
}
{\catcode`\&=4
\def\ampskip{&\omit\hfil&}
\catcode`\&=13
\let&0
\xdef\letampskip{\def&{\ampskip}}%
\gdef\letnovbamp{\let\novb&\let\tab&}
}
\def\begintable{
   \begingroup%
   \catcode`\|=13\letbartab\letvbbar%
   \catcode`\&=13\letampskip\letnovbamp%
   \def\multispan##1{
      \omit \mscount##1%
      \multiply\mscount\tw@\advance\mscount\m@ne%
      \loop\ifnum\mscount>\@ne \sp@n\repeat%
   }
   \def\|{%
      &\omit\widevline&%
   }%
   \ruledtable
}
\long\def\ruledtable#1\endtable{%
%
%
%
   \offinterlineskip
   \tabskip 0pt
   \def\widevline{\vrule width\thicksize}
   \def\endrow{\@mpersand\omit\hfil\crnorm\@mpersand}%
   \def\crthick{\@mpersand\crnorm\thickrule\@mpersand}%
   \def\crthickneg##1{\@mpersand\crnorm\thickrule
          \noalign{{\skip0=##1\vskip-\skip0}}\@mpersand}%
   \def\crnorule{\@mpersand\crnorm\@mpersand}%
   \def\crnoruleneg##1{\@mpersand\crnorm
          \noalign{{\skip0=##1\vskip-\skip0}}\@mpersand}%
   \let\nr=\crnorule
   \def\endtable{\@mpersand\crnorm\thickrule}%
   \let\crnorm=\cr
%
%
   \edef\cr{\@mpersand\crnorm\tablerule\@mpersand}%
   \def\crneg##1{\@mpersand\crnorm\tablerule
          \noalign{{\skip0=##1\vskip-\skip0}}\@mpersand}%
   \let\ctneg=\crthickneg
   \let\nrneg=\crnoruleneg
   \the\tableLETtokens
%
%
   \tabletokens={&#1}
%
%
   \countROWS\tabletokens\into\nrows%
   \countCOLS\tabletokens\into\ncols%
%
%
   \advance\ncols by -1%
   \divide\ncols by 2%
   \advance\nrows by 1%
%
%
   \iftableinfo %
      \immediate\write16{[Nrows=\the\nrows, Ncols=\the\ncols]}%
   \fi%
%
%
   \ifcentertables
      \ifhmode \par\fi
      \hbox to \hsize{
      \hss
   \else %
      \hbox{%
   \fi
      \vbox{%
         \makePREAMBLE{\the\ncols}
         \edef\next{\preamble}
         \let\preamble=\next
         \makeTABLE{\preamble}{\tabletokens}
      }
      \ifcentertables \hss}\else }\fi
   \endgroup
   \tablewidth=-\maxdimen
   \spreadwidth=-\maxdimen
}
\def\makeTABLE#1#2{
   {
   \let\ifmath0
   \let\header0
   \let\multispan0
%
%
   \ncase=0%
   \ifdim\tablewidth>-\maxdimen \ncase=1\fi%
   \ifdim\spreadwidth>-\maxdimen \ncase=2\fi%
   \relax
%
   \ifcase\ncase %
      \widthspec={}%
   \or %
      \widthspec=\expandafter{\expandafter t\expandafter o%
                 \the\tablewidth}%
   \else %
      \widthspec=\expandafter{\expandafter s\expandafter p\expandafter r%
                 \expandafter e\expandafter a\expandafter d%
                 \the\spreadwidth}%
   \fi %
   \xdef\next{
      \halign\the\widthspec{%
      #1
      \noalign{\hrule height\thicksize depth0pt}
      \the#2\endtable
%
      }
   }
   }
   \next
}
\def\makePREAMBLE#1{
   \ncols=#1
   \begingroup
   \let\ARGS=0
   \edef\xtp{\widevline\ARGS\tabskip\tabskipglue%
   &\ctr{\ARGS}\tstrut}
   \advance\ncols by -1
   \loop
      \ifnum\ncols>0 %
      \advance\ncols by -1%
      \edef\xtp{\xtp&\vrule width\thinsize\ARGS&\ctr{\ARGS}}%
   \repeat
   \xdef\preamble{\xtp&\widevline\ARGS\tabskip0pt%
   \crnorm}
   \endgroup
}
\def\countROWS#1\into#2{
   \let\countREGISTER=#2%
   \countREGISTER=0%
   \expandafter\ROWcount\the#1\endcount%
}%
\def\ROWcount{%
   \afterassignment\subROWcount\let\next= %
}%
\def\subROWcount{%
   \ifx\next\endcount %
      \let\next=\relax%
   \else%
      \ncase=0%
      \ifx\next\cr %
         \global\advance\countREGISTER by 1%
         \ncase=0%
      \fi%
      \ifx\next\endrow %
         \global\advance\countREGISTER by 1%
         \ncase=0%
      \fi%
      \ifx\next\crthick %
         \global\advance\countREGISTER by 1%
         \ncase=0%
      \fi%
      \ifx\next\crnorule %
         \global\advance\countREGISTER by 1%
         \ncase=0%
      \fi%
      \ifx\next\crthickneg %
         \global\advance\countREGISTER by 1%
         \ncase=0%
      \fi%
      \ifx\next\crnoruleneg %
         \global\advance\countREGISTER by 1%
         \ncase=0%
      \fi%
      \ifx\next\crneg %
         \global\advance\countREGISTER by 1%
         \ncase=0%
      \fi%
      \ifx\next\header %
         \ncase=1%
      \fi%
      \relax%
      \ifcase\ncase %
         \let\next\ROWcount%
      \or %
         \let\next\argROWskip%
      \else %
      \fi%
   \fi%
   \next%
}
\def\counthdROWS#1\into#2{%
\dvr{10}%
   \let\countREGISTER=#2%
   \countREGISTER=0%
\dvr{11}%
\dvr{13}%
   \expandafter\hdROWcount\the#1\endcount%
\dvr{12}%
}%
\def\hdROWcount{%
   \afterassignment\subhdROWcount\let\next= %
}%
\def\subhdROWcount{%
   \ifx\next\endcount %
      \let\next=\relax%
   \else%
      \ncase=0%
      \ifx\next\cr %
         \global\advance\countREGISTER by 1%
         \ncase=0%
      \fi%
      \ifx\next\endrow %
         \global\advance\countREGISTER by 1%
         \ncase=0%
      \fi%
      \ifx\next\crthick %
         \global\advance\countREGISTER by 1%
         \ncase=0%
      \fi%
      \ifx\next\crnorule %
         \global\advance\countREGISTER by 1%
         \ncase=0%
      \fi%
      \ifx\next\header %
         \ncase=1%
      \fi%
\relax%
      \ifcase\ncase %
         \let\next\hdROWcount%
      \or%
         \let\next\arghdROWskip%
      \else %
      \fi%
   \fi%
   \next%
}%
{\catcode`\|=13\letbartab
\gdef\countCOLS#1\into#2{%
   \let\countREGISTER=#2%
   \global\countREGISTER=0%
   \global\multispancount=0%
   \global\firstrowtrue
   \expandafter\COLcount\the#1\endcount%
   \global\advance\countREGISTER by 3%
   \global\advance\countREGISTER by -\multispancount
}%
\gdef\COLcount{%
   \afterassignment\subCOLcount\let\next= %
}%
{\catcode`\&=13%
\gdef\subCOLcount{%
   \ifx\next\endcount %
      \let\next=\relax%
   \else%
      \ncase=0%
      \iffirstrow
         \ifx\next& %
            \global\advance\countREGISTER by 2%
            \ncase=0%
         \fi%
         \ifx\next\span %
            \global\advance\countREGISTER by 1%
            \ncase=0%
         \fi%
         \ifx\next| %
            \global\advance\countREGISTER by 2%
            \ncase=0%
         \fi
         \ifx\next\|
            \global\advance\countREGISTER by 2%
            \ncase=0%
         \fi
         \ifx\next\multispan
            \ncase=1%
            \global\advance\multispancount by 1%
         \fi
         \ifx\next\header
            \ncase=2%
         \fi
         \ifx\next\cr       \global\firstrowfalse \fi
         \ifx\next\endrow   \global\firstrowfalse \fi
         \ifx\next\crthick  \global\firstrowfalse \fi
         \ifx\next\crnorule \global\firstrowfalse \fi
         \ifx\next\crnoruleneg \global\firstrowfalse \fi
         \ifx\next\crthickneg  \global\firstrowfalse \fi
         \ifx\next\crneg       \global\firstrowfalse \fi
      \fi
\relax
      \ifcase\ncase %
         \let\next\COLcount%
      \or %
         \let\next\spancount%
      \or %
         \let\next\argCOLskip%
      \else %
      \fi %
   \fi%
   \next%
}%
\gdef\argROWskip#1{%
   \let\next\ROWcount \next%
}
\gdef\arghdROWskip#1{%
   \let\next\ROWcount \next%
}
\gdef\argCOLskip#1{%
   \let\next\COLcount \next%
}
}
}
\def\spancount#1{
   \nspan=#1\multiply\nspan by 2\advance\nspan by -1%
   \global\advance \countREGISTER by \nspan
   \let\next\COLcount \next}%
\def\dvr#1{\relax}%
\def\header#1{%
\dvr{1}{\let\cr=\@mpersand%
\hdtks={#1}%
\counthdROWS\hdtks\into\hdrows%
\advance\hdrows by 1%
\ifnum\hdrows=0 \hdrows=1 \fi%
\dvr{5}\makehdPREAMBLE{\the\hdrows}%
\dvr{6}\getHDdimen{#1}%
{\parindent=0pt\hsize=\hdsize{\let\ifmath0%
\xdef\next{\valign{\headerpreamble #1\crnorm}}}\dvr{7}\next\dvr{8}%
}%
}\dvr{2}}
\def\makehdPREAMBLE#1{
\dvr{3}%
\hdrows=#1
{
\let\headerARGS=0%
\let\cr=\crnorm%
\edef\xtp{\vfil\hfil\hbox{\headerARGS}\hfil\vfil}%
\advance\hdrows by -1
\loop
\ifnum\hdrows>0%
\advance\hdrows by -1%
\edef\xtp{\xtp&\vfil\hfil\hbox{\headerARGS}\hfil\vfil}%
\repeat%
\xdef\headerpreamble{\xtp\crcr}%
}
\dvr{4}}
\def\getHDdimen#1{%
\hdsize=0pt%
\getsize#1\cr\end\cr%
}
\def\getsize#1\cr{%
\endsizefalse\savetks={#1}%
\expandafter\lookend\the\savetks\cr%
\relax \ifendsize \let\next\relax \else%
\setbox\hdbox=\hbox{#1}\newhdsize=1.0\wd\hdbox%
\ifdim\newhdsize>\hdsize \hdsize=\newhdsize \fi%
\let\next\getsize \fi%
\next%
}%
\def\lookend{\afterassignment\sublookend\let\looknext= }%
\def\sublookend{\relax%
\ifx\looknext\cr %
\let\looknext\relax \else %
   \relax
   \ifx\looknext\end \global\endsizetrue \fi%
   \let\looknext=\lookend%
    \fi \looknext%
}%
%
%
\def\tablelet#1{%
   \tableLETtokens=\expandafter{\the\tableLETtokens #1}%
}%
\catcode`\@=12
%

\title{ 
     Baryon stopping in high energy collisions and the extrapolation of
     hadron production models to Cosmic Ray energies
}

 \author{  J.~Ranft }
  \address{
 Physics Dept. Universit\"at Siegen, D--57068 Siegen, Germany,
 e--mail: Johannes.Ranft@cern.ch}
\date{\today}


\maketitle

\vspace{3mm}
\begin{abstract}

 A new striking feature of hadron production in  nuclear
 collisions is the large stopping of the
 participating nucleons 
 in hadron--nucleus and nucleus--nucleus collisions. This enhanced
 baryon stopping can be understood introducing new diquark breaking
 mechanisms in multistring models of hadron production. Here we show,
 that similar diquark breaking mechanisms occur at high energy even in
 hadron--hadron collisions. This effect leads to significant changes in
 the extrapolation of these models to Cosmic Ray energies. 

\end{abstract}

\vskip 10mm 

\section{Introduction}

 A new  feature of hadron production in  nuclear
 collisions is the large stopping of the
 participating nucleons 
 in hadron--nucleus and nucleus--nucleus collisions. Experimental data
 demonstrating this effect have been presented in
 \cite{NA35FIN,Alber98}. 

Multistring fragmentation models like the Dual Parton Model
 (DPM) \cite{Capella79,capdpm,Engelthesis} 
or similar models
did originally not show this enhanced stopping in nuclear collisions.
Therefore, in order to incorporate the effect into multistring
fragmentation models
 new diquark breaking DPM--diagrams acting in
 hadron--nucleus and nucleus--nucleus collisions were 
 proposed by Capella and Kopeliovich
 \cite{Capella96} and investigated in detail by Capella and
 Collaborators \cite{Capella99a,Capella99,Capella99b}. 
 Similar ideas were discussed by Vance and
 Gyulassy\cite{Vance99}
   and by Casado \cite{Casado99}.
 
In the present paper we desribe  a new diquark breaking diagram, which
acts in nuclear collisions as well as in hadron--hadron collisions at
sufficiently high energy. 
The new diagram becomes important only at energies well
above the energies of the CERN--SPS heavy ion experiments which have
been discussed in the papers quoted above. The effect of this new
diagram modifies significantly the extrapolation of multistring
fragmentation models to Cosmic Ray energies. This is our main interest
in the present paper.

The diquark breaking diagrams investigated by Capella and 
Collaborators \cite{Capella99a,Capella99}
use sea quarks, which in nuclear collisions according to the
Glauber model are needed to implement the multiple collisions in
the nucleus. In the Dual Parton Model in the collider energy
range there are further multiple collisions, again implemented
using sea quarks at the ends of multiple chains, which occur due
to the unitarization procedure. These are the sea quarks which provide
the diquark breaking in our new diagram.

To get definitive predictions from the old and the new diquark breaking
diagrams, we introduce the diagrams into the Monte Carlo  version of a
multistring fragmentation model.
DPMJET simulates  hadron production in the framework
of the Dual Parton Model with emphasis  as well in the central
as in the fragmentation
region. 

Previous versions of  the DPMJET
event generator  were described in detail  
 \cite{DPMJETII,Ranftsare95,Ranft95b,dpmjet2324}.
Here we use the present version DPMJET--II.5 
\cite{Ranft99a,Ranft99b} of the code.

The most important applications of DPMJET so far were
 in Collaboration with Battistoni,  Forti and others
\cite{Battistoni95a,Battistoni95b,Battistoni96a,Battistoni99a} 
for the simulation of the Cosmic Ray cascade in the
HEMAS--DPM code system.

Observations
like rapidity plateaus and average transverse momenta 
rising with energy,
KNO scaling violation, 
transverse momentum--multiplicity correlations and
{\it minijets} pointed out, 
that soft and hard processes are closely related.
These properties were understood within the two--component Dual Parton
Model
\cite{CTK87,DTUJETPR92,DTUJETZP91,DTUJET92a,DTUJET92b,DTUJETDIFF,DTUJET93}. 
The hard component is introduced applying lowest order
of perturbative hard constituent scattering \cite{CKR77}. 
Single diffraction dissociation
is represented by a triple--pomeron
exchange (high mass single diffraction) and a low mass component.

The Dual Parton model provides a framework not only for the
study of hadron--hadron interactions, but also for the
description of particle production in hadron--nucleus and
nucleus--nucleus collisions at high energies. Within this model
the high energy projectile undergoes a multiple scattering as
formulated in Glaubers approach; particle production is 
realized by the fragmentation of colorless parton--parton chains
constructed from the quark content of the 
interacting hadrons and nuclei.

In Section II we describe the new diagrams introduced in multistring
models like the DPM in order to get a better description of baryon
stopping.
In Section III we compare the model to data and in Section IV we
discuss the properties of the model at the highest energies. 
%

%

\section{Implementation of new DPM diagrams for an improved
description of baryon stopping in ihadron--hadron, 
hadron--nucleus and nucleus--nucleus
collisions}

\subsection{Diquark fragmentation, the popcorn mechanism}

The fragmentation of diquarks is slightly more complicated than
the fragmentation of a quark jet.
As justified by  Rossi and Veneziano \cite{Rossi97} the baryon can
be pictured as made out of three quarks bound together by three
strings which join in a so called string junction point.
In diagrams we can characterize the baryons 

(i)by the three quarks
and the string junction or 

(ii)by a quark and a diquark, in this case the string junction always
goes with the diquark.

In all the diagrams discussed in
this section we will either plot the quark and the diquark or, if the
diquark breaks, the three quarks and the string junction. The quarks and
the two quarks of a diquark are plotted as solid lines, the string
junction is plotted as a dashed line.

\begin{figure}[thb] \centering
\begin{turn}{-90}
 \psfig{figure=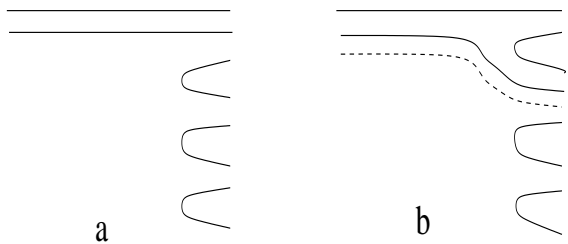,width=4.0cm,height=8.0cm}
\end{turn}
 \caption{{\bf (a) }Conventional diquark fragmentation, the baryon is
 produced in the first fragmentation step. {\bf (b) }The popcorn
 mechanism: a meson is produced in the first fragmentation step, the
 baryon appears in the second  or later fragmentation
 steps. 
 We plot the lines for the two quarks of the diquark as solid lines,
  the string junction is plotted as a dashed line.
 \label{popc}
 }
 \end{figure}

There are 
 two possibilities for the first fragmentation step of a diquark, see
 Fig.\ref{popc}.
 Either we
get in the first step a baryon, which contains
both quarks of the diquark and the string junction or we get  in the
first step a meson containing only one of the two quarks and the
baryon is produced in one of  
the following fragmentation step. This
mechanism is well known, it is presented in the review on the
Dual Parton Model
\cite{capdpm} and it was investigated 
for instance in \cite{SUKHA,KOPEL}.
 This mechanism was implemented
from the beginning in the 
BAMJET--fragmentation code \cite{BAMJET1,BAMJET}
used previously in DPMJET. 
This mechanism is also implemented under the name {\it popcorn}
fragmentation in the Lund chain fragmentation model JETSET
\cite{JETSET,AND85} which is presently used in DPMJET. 
\\
 What happens in the model with the popcorn mechanism compared to the
model without can be most easily seen looking at the proton
rapidity distribution in p--p collisions.
The two maxima in the target and projectile
fragmentation region of the proton rapidity distribution shift by
about half a unity to the center,  these maxima become
wider and  correspondingly the dip in the center is reduced.
At the same time the Feynman $x$ distributions of mesons get
a component at larger Feynman $x$. The effects in hadron--nucleus and
nucleus--nucleus collisions are quite similar.

The popcorn effect as implemented in the JETSET fragmentation code
\cite{JETSET} determines  not only the fragmentation of a diquark as
plottet in Fig. \ref{popc}. A second effect of the popcorn mechanism is
in baryon pair production inside the fragmenting chain. Here the popcorm
mechanism determines the ratio of $B-\bar B$ ( baryon and
antibaryon are neighbors in the chain) to $B-M-\bar B$ ( a meson is
produced between the baryon
and the antibaryon) production.

The popcorn mechanism  alone is  not enough to explain the baryon
stopping observed experimentally in hadron--nucleus and nucleus--nucleus
collisions
 \cite{NA35FIN,Alber98}, this will be discussed in more detail  in
 Section III.

\subsection{New diquark breaking DPM diagrams 
in hadron--nucleus and nucleus--nucleus
collisions}

\begin{figure}[thb] \centering
\begin{turn}{-90}
 \psfig{figure=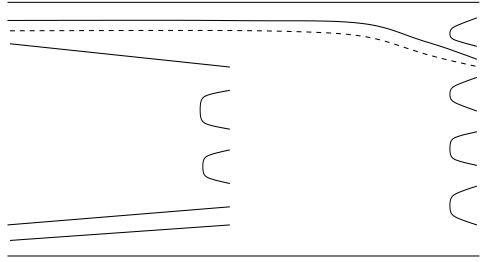,width=4.0cm,height=7.0cm}
\end{turn}
 \vspace*{2mm}
 \caption{The first C--K diquark breaking mechanism \protect\cite{Capella96}
 plotted for a nucleon--nucleon collision.
 \protect\label{diqb}
 }
 \end{figure}

 New diquark breaking DPM--diagrams mainly of interest in
 hadron--nucleus and nucleus--nucleus collisions were 
 proposed by Capella and Kopeliovich
 \cite{Capella96} and investigated in detail by Capella and
 Collaborators \cite{Capella99a,Capella99,Capella99b}. 
 Similar ideas were discussed by Vance and
 Gyulassy\cite{Vance99}.
Capella and Kopeliovich\cite{Capella96} did discuss 
in detail their first diquark
breaking mechanism (see Fig.\ref{diqb}, where this mechanism is
characterized for nucleon--nucleon collisions), in this case the
valence--diquark breaks into the two quarks, the baryon is produced in
the second or in later fragmentation steps. 

Looking only at  the diagrams for a hadron--hadron collision, 
one does not see any difference between the popcorn
 mechanism and the first  Capella--Kopeliovich (C--K) diquark breaking
 mechanism.
 However, the mechanism of this first  
 (C--K) diquark breaking
 mechanism  differs in detail and especially in hadron--nucleus and
 nucleus--nucleus collisions from the popcorn mechanism discussed
 above.
 
 Capella and Kopeliovich 
 \cite{Capella96} write for the baryon rapidity distribution in the
 diquark breaking (DB) component  
\begin{equation}
\frac{dN_{DB}}{dy}(y)\approx [\rho_{q_v}(y) + \rho_{q_v}(-y)]
\end{equation}
where $\rho_{q_v}$ is the valence quark rapidity distribution.
In hadron--nucleus (NA) collisions  we have
\begin{equation}
\sigma^{NA}_{in}=\sigma^{NA}_{DP} + \sigma^{NA}_{DB}
\end{equation}
where DP stands for diquark preserving. Studying the A behaviour of
$\sigma^{NA}_{DB}$ they find, that in N--A collisions
$\sigma^{NA}_{DB}$ rises faster with
A than the diquark preserving component $\sigma^{NA}_{DP}$. The same
happens in nucleus--nucleus collisions.
 
 Nevertheless, despite these important 
 differences between the popcorn effect and
 the first C--K mechanism we find that implementing the first 
 C--K mechanism in DPMJET did not give
 any new feature of baryon stopping, which could not also 
 be obtained from
 the popcorn mechanism. Therefore, 
 we continue to use in DPMJET--II.5 the
 popcorn mechanism (which as discussed above also acts on
 baryon--antibaryon pair production in the middle of a chain)
 instead of the C--K mechanism, so far we did not find any argument  to
 use both effects.

\begin{figure}[thb] \centering
\begin{turn}{-90}
 \psfig{figure=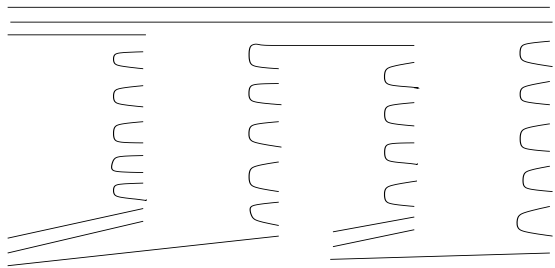,width=4.0cm,height=8.0cm}
\end{turn}
 \vspace*{2mm}
 \caption{The diquark--conserving diagram for a nucleon--nucleus
 collision with two participants of the target nucleus.
 \protect\label{diqc}
 }
 \end{figure}

 More interesting is the second C--K mechanism, which was proposed but 
 not discussed in detail in \cite{Capella96}. This mechanism was 
 discussed in detail by Capella and 
 Collaborators\cite{Capella99a,Capella99,Capella99b}. In
 Fig.\ref{diqc} we plot first the diquark--conserving diagram for a
 nucleon--nucleus collisions with two participants of the target
 nucleus. This is the traditional way for such a collision in the DPM.

\begin{figure}[thb] \centering
\begin{turn}{-90}
 \psfig{figure=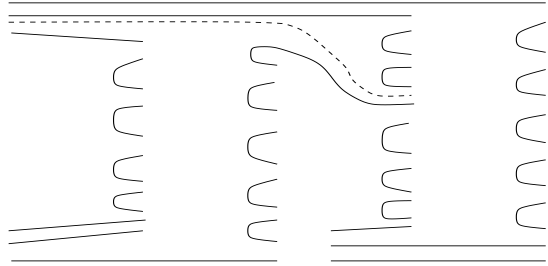,width=4.0cm,height=8.0cm}
\end{turn}
 \vspace*{2mm}
 \caption{The Glauber sea quark mechanism of baryon stopping GSQBS for a
 nucleon--nucleus collision with two 
 participants of the target nucleus. This
 is the second C--K diquark--breaking mechanism \protect\cite{Capella96}.
 \protect\label{diqsea}
 }
 \end{figure}

In Fig.\ref{diqsea} we plot the second C--K diquark--breaking diagram
for the same collision. 
Now the second 
valence quark from the broken diquark 
in the first C--K diagram 
in Fig.\ref{diqb} is
replaced by a Glauber 
sea quark from the nucleon projectile. Therefore, we will
call the mechanism the Glauber 
sea quark mechanism of baryon stopping GSQBS. The probability of such a
diquark splitting rises if the considered nucleon is involved in more
than two interactions.

Capella et al. \cite{Capella99a} write for the rapidity
distribution of the net baryon production $\Delta B = B -\bar B $
in A--A collisions
\begin{equation}
\frac{dN_{AA\rightarrow \Delta B}}{dy}(y)=\frac{\bar n_A}{\bar n}[\bar
n_A(\frac{dN^{\Delta B}_{DP}}{dy}(y))_{\bar n/\bar n_A } + (\bar
n-\bar n_A)(\frac{dN^{\Delta B}_{DB}}{dy}(y))_{\bar n/\bar n_A}] 
\end{equation}
where $\bar n_A$ is the average number of participants in each
nucleus and $\bar n $ the average number of collisions.  For $\bar
n = \bar n_A$ only the diquark preserving component is present.
$\frac{dN^{\Delta B}_{DB}}{dy}$ is the rapidity density of the
diquark breaking component which behaves approximately like
$\exp [-\frac{1}{2}|y_{\Delta B} -y_{Max}|]$.

The GSQBS has been
implemented in DPMJET--II.5 and we will see in Section III, that in
nucleon--nucleus collisions and nucleus--nucleus collisions we are able
with this mechanism 
to fill the dip in the baryon rapidity distributions at central rapidity
in agreement to the experimental data. As discussed already in detail in
\cite{Capella99,Capella99b} 
this mechanism also contributes to increase the Hyperon
production in nucleon--nucleus and nucleus--nucleus collisions.

\subsection{The Casado diagram 
}

\begin{figure}[thb] \centering
\begin{turn}{-90}
 \psfig{figure=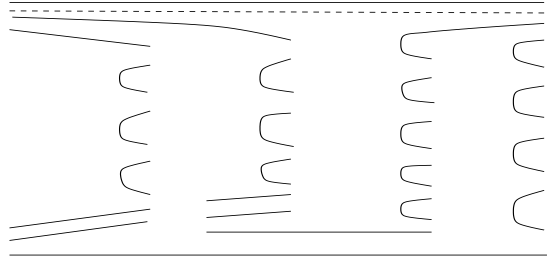,width=4.0cm,height=8.0cm}
\end{turn}
 \vspace*{2mm}
 \caption{The diagram introduced by Casado \protect\cite{Casado99}
 for a nucleon--nucleus
 collision with two participants of the target nucleus. The diquark
 contains one Glauber sea quark.
 \protect\label{casado}
 }
 \end{figure}

A  new diagram was also introduced by Casado \cite{Casado99}. The
diagram is plotted  for a nucleon--nucleus collision with two
participants of the target nucleus
in Fig. \ref{casado}. Here the fragmenting diquark contains a Glauber
sea quark and hadronizes like a valence diquark producing baryons mainly
in the fragmentation region. However, the flavor content of the baryon 
is changed. This diagram gives another contribution to hyperon
production in nuclear collisions. We have implemented the Casado diagram
in DPMJET--II.5 and it is used in addition to the GSQBS diagram for all
nuclear collisions compared in  Section  III to data.

\subsection{The new diquark breaking diagram in hadron--hadron and
nuclear collisions at high energy.
A new extrapolations of DPMJET--II.5 to Cosmic Ray energies.
}

This new diagram follows necessarily from the GSQBS diagram if
we go to high energy.
At high energies we have multiple collisions even in hadron--hadron
collisions due to the unitarization procedure. We call the sea quarks at
the ends of the additional chains in 
this case {\it unitary sea quarks}. The
Glauber sea quarks are needed in nuclear collisions already at rather
low energies, for instance at the energies of heavy ion collisions at
the CERN--SPS. In contrast to this, unitary sea quarks appear in
significant numbers in
hadron--hadron and nuclear collisions only at rather high energies, for
instance at the energies of the CERN--SPS collider or the TEVATRON
collider, they are important in the Cosmic Ray energy region.

With the unitary sea
quarks at the ends of the chains from the secondary collisions we obtain
a new mechanism for baryon stopping, which will become effective at very
high energies.

\begin{figure}[thb] \centering
\begin{turn}{-90}
 \psfig{figure=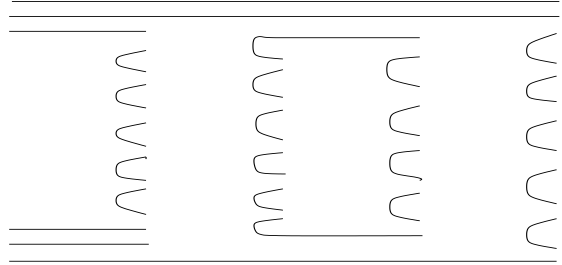,width=4.0cm,height=8.0cm}
\end{turn}
 \vspace*{2mm}
 \caption{Standard DPM diagram for a nucleon--nucleon interaction with
 one additional soft secondary interaction 
 induced by the unitarzation procedure.
 There is one valence--valence and one sea--sea interaction, each
 represented by a pair of chains.
 \protect\label{normal2}
 }
 \end{figure}

\begin{figure}[thb] \centering
\begin{turn}{-90}
 \psfig{figure=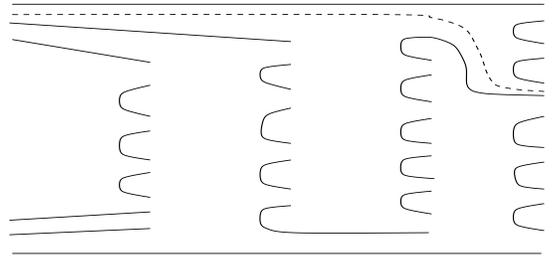,width=4.0cm,height=8.0cm}
\end{turn}
 \vspace*{2mm}
 \caption{New DPM diagram for a nucleon--nucleon interaction with one
 additional soft secondary interaction 
 induced by the unitarzation procedure. The
 diquark is split and an unitarity sea quark is used similar to Fig.
 \protect\ref{diqsea} to shift the baryon in one of the chains. We call this the
 unitary sea quark mechanism for baryon stopping USQBS.
 \protect\label{new2}
 }
 \end{figure}

In Fig.\ref{normal2} we plot the standard DPM diagram for a
nucleon--nucleon interaction with two soft  interactions induced
by the unitarization procedure. There is one valence--valence and one
sea--sea interaction, each
represented by a pair of chains. In analogy to Fig. \ref{diqsea} we
construct from this the new diagram for baryon stopping in Fig.
\ref{new2}.
  The
 diquark is split and an unitary sea quark is used 
 to have the  baryon only in the second or later fragmentation steps 
 in one of the chains. We call this the
 unitary sea quark mechanism for baryon stopping USQBS. Also here the
 probability for such a diquark splitting rises if there are more than 2
 interactions of the hadrons involved. Obviously, 
 this mechanism leads to  Feynman $x$ distributions of baryons in
 p--p collisions becoming  softer and Feynman $x$ distributions of
 mesons becoming  harder than without the USQBS mechanism.
 
One word about the implementation of the diquark breaking
diagrams in DPMJET--II.5. In a Monte Carlo version of the Dual
Parton Model we do of course not use the parametrizations used
by Capella et al.\cite{Capella96,Capella99a,Capella99,Capella99b}.
In each event DPMJET first constructs the conventionel chains
without diquark breaking. From this before implementing the
chain fragmentation we construct the chain structure of the
event including
the diquark breaking mechanisms.
In each elementary collision in DPMJET the number of
participating Glauber sea quarks and unitary sea quarks is well
known and given a probability for the diquark breaking diagram,
each possible sea quark gets the chance to get involved in the
diquark breaking mechanism. Also in DPMJET we do not need any
parametrization for the rapidity distributions of the DP and DB
components, with given $x$--values of the participating
diquarks, valence quarks and sea quarks and using the JETSET
chain fragmentation we get the resulting hadrons at the correct
positions, energy conservation in the event is satisfied in this
way. If the diquark is broken, we have to convert the $x$ of the
diquark into the two $x$--values of the two valence quarks.
At low energy and for original diquark--quark chains of low
invariant mass we have to reject the diquark breaking at times.
If the diquark breaking is kinematically impossible, we retain
the original DP mechanism.

 In contrast to the  GSQBS mechanism, which leads already
 to effects in nuclear collisions 
 at the energy of the CERN--SPS, we have at present
 no data available to prove that this USQBS mechanism is a valid
 extension of the DPM. There is no data for baryon stopping in
 proton--proton or antiproton--proton collisions at collider energies.
 Unfortunately, the fragmentation region at large Feynman $x$ has not
 been investigated experimentally with enough detail 
 at any of the hadron--hadron
 colliders.
However, if the GSQBS mechanism is the correct mechanism responsable for
the baryon stopping effects found in nuclear collisions, then also the
USQBS mechanism should modify the collisions at collider energies and
beyond. In  Section IV, where we present the properties of DPMJET--II.5
at Cosmic Ray energies we present always two extrapolations of the
model. In one version (shortly characterized as version 55 in the plots)
we have both the GSQBS and USQBS mechanism, in a second version
(characterized in the plots as
version 50) we have only the GSQBS mechanism. Only collider experiments
on baryon stopping or Cosmic Ray observations can prove, which version
is the better one. From theoretical prejudices however there would be
the claim, that the version with both the GSQBS and USQBS mechanisms
is to be prefered.

 The version of the model with both GSQBS and USQBS (designated by 55 in
 the plots) is the new version
 of the model with all new baryon stopping effects.  The model with only
 GSQBS (designated by 50 in the plots) 
 serves as reference only, it corresponds in its high energy behaviour 
 closely to previous
 versions of DPMJET or to other models without the new baryon stopping
 mechanisms.

\subsection{New parameters connected with the diquark breaking
diagrams
}

For each of the new diquark breaking diagrams described in this Section
we have to introduce a new parameter. These parameters give the
probability for the diquark breaking mechanisms to occur, 
given a suitable sea
quark is available and given that the diquark breaking mechanism is
kinematically allowed. For an original diquark--quark chain of small
invariant mass, which originally 
just fragments into two hadrons, the diquark
breaking is often not allowed at small energies.

The optimum values of the new parameters are determined by comparing
DPMJET--II.5 with experimental data on net--baryon distibutions. We use
 for the GSQBS and USQBS mechanisms the default parameters 0.45 and for
 the Casado diagram the default parameter 0.5. 

\subsection{Production of strange particles}

Capella et al \cite{Capella99a,Capella99,Capella99b} discuss the new
diagrams for baryon stopping also with the  interest in strangeness
production in nuclear collision. They discuss in detail the production
of hyperons and antihyperons. We will not repeat their discussion here,
we only shortly summarize the different mechanisms, which contribute to
strangeness production.

Studies of strangeness production within
DPMJET were  given in \cite{MRCT93,CTR1,CTR94}.
Enhanced generation of strange particles,
in particular of
strange antibaryons, has been proposed
as a signal for the formation of quark gluon plasma in dense
hadronic matter~\cite{rafel,rafel1}.
Recent data from experiments at the CERN SPS have already
been interpreted
within this scheme.
However, we find it worthwhile to pursue the study of
tested conventional models
without QGP formation like the DPM before drawing final
conclusions.
The DPM  is an independent string model.
 Since the individual strings are universal building
blocks of the model, the ratio of {\it produced} strange particles
over non--strange ones will be approximately the same in all
 reactions. However, since some strings contain sea quarks at one
or both ends and since strange quarks are present in the proton sea,
it is clear, that, by increasing the number of those strings, the
ratio of strange over non--strange particles will increase.
This will be the case for instance, when increasing the
centrality in a nucleus--nucleus collision. It is obvious, that
the numerical importance of the effect will depend on the assumed
fraction of strange over non--strange quarks in the proton sea.
 The rather extreme case leading to a maximum increase of strangeness
would be to assume a SU(3) 
symmetric sea (equal numbers of $u$, $d$ and $s$
flavors).
 We express the
amount of SU(3) symmetry of the sea chain ends by our parameter
$s^{sea}$ defined as
$s^{sea} =  {2<s_s>}/({<u_s>+<d_s>})$
where the $<q_s>$ give the average numbers of sea quarks at the
sea chain ends. Usually, DPMJET uses  the default value $s^{sea}$ = 0.5.
\\
There are a number of effects, which change the number of strange
hadrons, especially strange and multistrange baryons 
and antibaryons in the model: 

(i)The presence of sea $qq$ and $\bar q \bar q$ diquarks at sea--chain
ends \cite{Ranft93a,Moehring93a}.

(ii)The popcorn effect.

(iii)The diquark breaking diagrams discussed in Section II.C, see also
 \cite{Capella99a,Capella99,Capella99b}. 

(iv)The secondary interactions of co--moving produced particles
 \cite{Capella95,Capella95b}, see also Section II.G.

(v)Effects of string fusion \cite{Merino,Moehring93b}
and percolation of strings \cite{Armesto96,Braun97}.

It is certainly beyond the scope of the present paper, to discuss all of
these effects in detail. The effects (i) to (iv) are contained in
DPMJET--II.5.

\subsection{Final state interactions of co--moving secondaries in
nuclear collisions}

Analyzing the rapidity distributions of produced $\Lambda$ and $\bar
\Lambda$ perticles in central heavy ion collisions at CERN--SPS energies
within the Dual Parton Model Capella et al \cite{Capella95,Capella95b}
noted the need for secondary interactions of co--moving secondaries to
understand the data. Secondary interactions of produced hadrons are also
considered in other Monte Carlo models  for hadron production in nuclear
collisions \cite{VENUS,Bass98,LUCIAE}.

 Capella et al \cite{Capella95,Capella95b}
 only introduced the following secondary
$\pi + N \rightarrow K + \Lambda$
and
$\pi + \bar N \rightarrow K + \bar \Lambda$.
A reasonable cross section for these reactions 
of $\sigma \approx$ 1.5 mb was needed in \cite{Capella95,Capella95b}
to understand the data. 
 All of
this was done in  \cite{Capella95,Capella95b} using analytical methods,
not a Monte Carlo event generator. 
Final state interactions of co--moving secondaries was also used in
\cite{Armesto97,Armesto99} to explain the $J/\psi$ supression in Pb--Pb
collisions.

The method of Capella et al  \cite{Capella95,Capella95b} was implemented
in DPMJET already in 1995. The only reason to discuss final
state interactions of co--moving secondaries here is our
comparison in the following Section of DPMJET--II.5 in this version
with experimental data.
In DPMJET the method is used to modify the
Monte Carlo events 
and we use only a cross section of
$\sigma \approx$ 1.0 mb. 
First results reproducing essentially the
results of  \cite{Capella95,Capella95b} are described shortly in
\cite{Ranftsare95} and in more detail in a unpublished code write--up
\cite{Ranft95b}.  We would like to stress, this method as implemented 
 with only the two reactions given above, can only be
considered as a first preliminary step. A better method of final state
interactions should be implemented as a Monte Carlo method from the
beginning and  more types of 
secondary interactions should be taken into account
including the reverse reactions.
With the present method we can not expect to obtain reasonable results
in situations with much higher secondary particle densities than in the
heavy ion collisions at the CERN--SPS energy. Therefore, it is not
recommended to use the secondary interaction option of DPMJET at RHIC
or CERN--LHC energies.



\section{Comparing DPMJET--II.5 to data with emphasis to baryon
stopping}


\subsection{Comparing to data  on leading particle production 
in hadron--hadron collisions}

 The leading particle
production is very important for the Cosmic Ray cascade simulation.
Also,
the leading baryons are directly influenced by the diquark breaking
mechanisms.

\begin{figure}[thb]
\begin{center}
 \psfig{figure=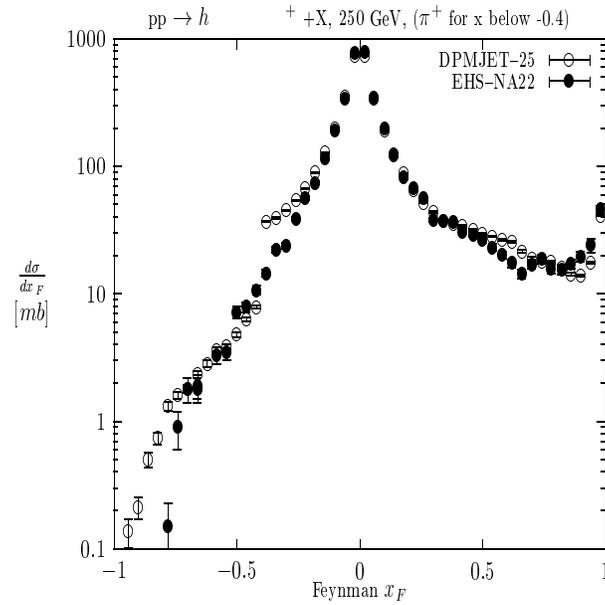,width=8cm,height=8cm} 
\end{center}
\vspace*{-3mm}
\caption{Comparison of Feynman--x distributions of positively charged
hadrons
 produced in proton--proton collisions at 250 GeV. The
experimental data are from the EHS--NA22 Collaboration
\protect\cite{Adamus88a}.
\protect\label{pp250hplux}
}
\end{figure}

In Fig.  \ref{pp250hplux}   
we compare the model with NA-22 data on
the Feynman--$x$ distribution of positively charged hadrons  
 produced in 250 GeV
$pp$ collisions. In Fig. \ref{pp250hplux} we observe in particular
(compared to previous DPMJET versions)  a
better agreement of the model predictions with the forward production of
protons.  

\begin{figure}[thb] \centering
 \psfig{figure=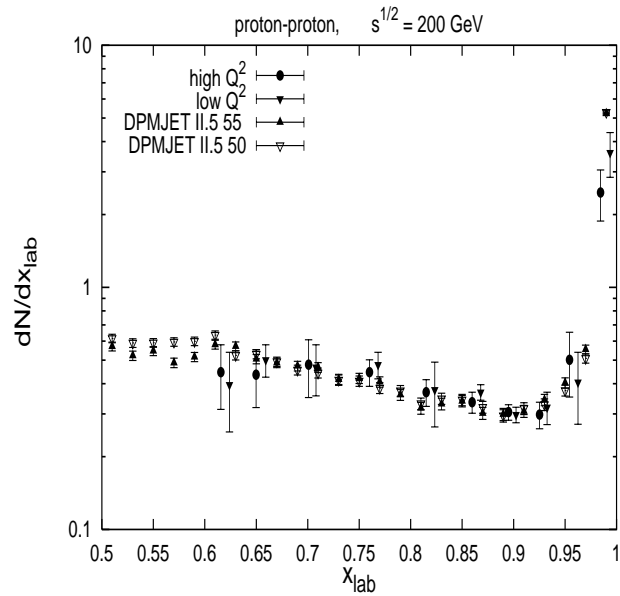,width=8.0cm,height=8.0cm}
\vspace*{8mm}
\caption{
Energy fraction \protect$x_{lab}$ carried by the leading proton. The data are
photoproduction and DIS measurements at $\sqrt s$ = 200 GeV
\protect\cite{Garfagnini98,Schmidke98} 
shown as symbols compared to both versions of DPMJET--II.5 for p--p
collisions at  $\sqrt s$ = 200 GeV. 
\protect\label{lead-pro3}
}
\end{figure}

Distributions of leading protons are compared more directly 
to data in   Fig.  \ref{lead-pro3}. The  Figure is adopted 
from  talks of Engel \cite{Engel99a,Engel99b}. But of course, we present
now the DPMJET--II.5 results.
 In Fig.  \ref{lead-pro3} we compare 
the distribution in the energy fraction $x_{lab}$ 
carried by the leading proton. 
The data are
photoproduction and DIS measurements from the HERA Collider 
at $\sqrt s$ = 200 GeV
\cite{Garfagnini98,Schmidke98}. We compare to DPMJET--II.5 for p--p
collisions at  $\sqrt s$ = 200 GeV. The forward production of leading
protons is not expected to depend strongly on the reaction channel. It
is found, that DPMJET--II.5 agrees much better to the data than
older DPMJET versions, see \cite{Engel99b}. This
comparison demonstrates, that the leading proton distribution at $\sqrt
s$ = 200 GeV is still rather flat.
As shown in Fig.  \ref{lead-pro3} at $\sqrt s$ = 200 GeV  
both versions of DPMJET--II.5 (discussed in Section II)
with GSQBS and USQBS (designated in the plot by 55) and with only GSQBS
(designated in the plot by 50), 
lead still to very similar $x_{lab}$ distributions of
secondary protons. Unfortunately therefore, these data cannot be used to
discriminate between the two DPMJET--II.5 versions.

\begin{figure}[thb] \centering
 \psfig{figure=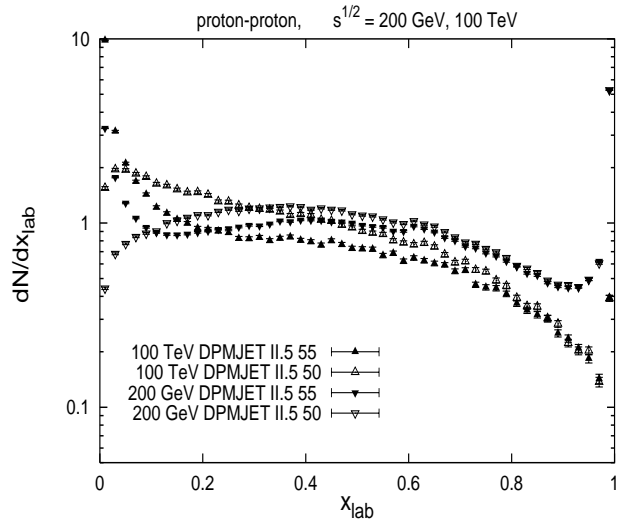,width=8.0cm,height=7.0cm}
\vspace*{8mm}
\caption{
Energy fraction \protect$x_{lab}$ carried by the leading baryon. 
 at $\sqrt s$ =200 GeV and 100 TeV
according to both versions of  DPMJET--II.5 for p--p
collisions. 
\protect\label{lead-bar100}
}
\end{figure}

 In Fig. \ref{lead-bar100} we compare the leading baryon distribution in
 the energy fraction  $x_{lab}$ according to both versions 
 of DPMJET--II.5 at $\sqrt s$ = 200 GeV and 100 TeV. At  $\sqrt s$ = 200
 GeV we find at  $x_{lab}$ obove 0.5 like for the leading proton
 distribution in Fig.  \ref{lead-pro3} hardly any difference between
 both versions of the model. At  $\sqrt s$ =  100 TeV the differences
 between both versions are significant even for  $x_{lab}$ above 0.5. At
 smaller  $x_{lab}$ values, say below  $x_{lab}$ =0.2 we find
 significant differences between both versions at both energies. The
 $x_{lab}$ distribution of leading protons looks rather similar to Fig.
 \ref{lead-bar100}. We  conclude leading proton or leading baryon
 distributions in the energy fraction  $x_{lab}$ below  $x_{lab}$ =0.5
 could be quite usefull to demonstrate the effect of the USQBS diagrams
 in  experimental data.

%
%
%
%

 
\subsection{Comparing to data in hadron--nucleus and
nucleus--nucleus collisions}

We now turn to collisions with nuclei. We want to compare the
model with the diquark breaking mechanisms to data in hadron--nucleus
and nucleus--nucleus collisions. Therefore we have to demonstrate thet
the model reproduces correctly the global properties of these nuclear
collisions. More comparisons are presented in \cite{Ranft99a}.

In Table 1 we compare average 
multiplicities of negatively charged hadrons calculated with
DPMJET--II.5 in
hadron--hadron, minimum bias hadron--nucleus and central
nucleus--nucleus collisions with experimental data.

{\bf Table 1.}
Comparison of average multiplicities of produced negatively charged 
hadrons in
proton--proton, minimum bias proton--nucleus and central
nucleus--nucleus collisions at 200 GeV.
\vskip 5mm
\begintable
Collision   | DPMJET--II.5 | Exp.|Reference ~\cr
 p--p     | 2.85  | 2.85 $\pm$ 0.03 | \cite{marek}  \cr
 p--S         | 5.10  |5.0  $\pm$0.2 | \cite{NA35}  \cr
 p--Ar         |5.30   |5.39  $\pm$0.17 | \cite{DeMarzo82}  \cr
 p--Ag         | 6.18  |6.2  $\pm$0.2 | \cite{Brick89}  \cr
 p--Xe         | 6.43  |6.84  $\pm$0.13 | \cite{DeMarzo82}  \cr
 p--Au         | 6.81  |7.0  $\pm$0.4 | \cite{Brick89}  \cr
 p--Au         | 6.81  |7.3  $\pm$0.3 | \cite{Baechler91}  \cr
 S--S central         |103   |98  $\pm$3 | \cite{Alber98}  \cr
 S--Ag central         |174   |186  $\pm$11 | \cite{Alber98}  \cr
 S--Au central         | 202  |225  $\pm$12 | \cite{Alber98}  \endtable
 \clearpage
{\bf Table 2.}
Comparison of average multiplicities of produced strange 
hadrons in
 central S--S and S--Ag
 collisions at 200 GeV.
 For DPMJET--II.5 we give  the results for the model  with
 secondary interactions  of comovers 
 and it is assumed that $\Sigma^0$ and
 $\bar \Sigma^0$ have decayed. The experimental data are from the NA35
 Collaboration \cite{Seyboth97}.
\vskip 5mm
\begintable
Reaction|Particle | DPMJET--II.5 with sec.int.| Exp.\cite{Seyboth97} ~\cr
S--S | $\Lambda$  | 7.3 |9.4 $\pm$ 1.0   \cr
S--S | $\bar \Lambda$  |1.22 |2.6 $\pm$ 0.3   \cr
S--S |$K^0_S$  |9.1 |10.5 $\pm$ 1.7   \cr
S--S |$K^+$  | 11.4 |12.5 $\pm$ 0.4   \cr
S--S |$K^-$  | 6.9 |6.9 $\pm$ 0.4   \cr
S--Ag | $\Lambda$  |13.1 |15.2 $\pm$1.2    \cr
S--Ag | $\bar \Lambda$  | 2.0 |2.6 $\pm$ 0.3   \cr
S--Ag | $K^0_S$  | 15.5 |15.5 $\pm$1.5    \cr
S--Ag | $K^+$  | 19.7 |17.4 $\pm$ 1.0   \cr
S--Ag | $K^-$  | 11.4 |9.6 $\pm$ 1.0   \endtable
\vskip 10mm 


In Table 2  we give the  average multiplicities 
of produced strange hadrons in central S--S and S--Ag 
 collisions at 200 GeV.
 We give the  DPMJET--II.5  
 results for the model  with
 secondary interactions and it is assumed, that $\Sigma^0$ and
 $\bar \Sigma^0$ have decayed. The experimental data for central S--S
and  S--Ag collisions are from the NA35
 Collaboration \cite{Seyboth97}. 
 
 
\subsection{Baryon stopping}

\begin{figure}[thb]
\begin{center}
 \psfig{figure=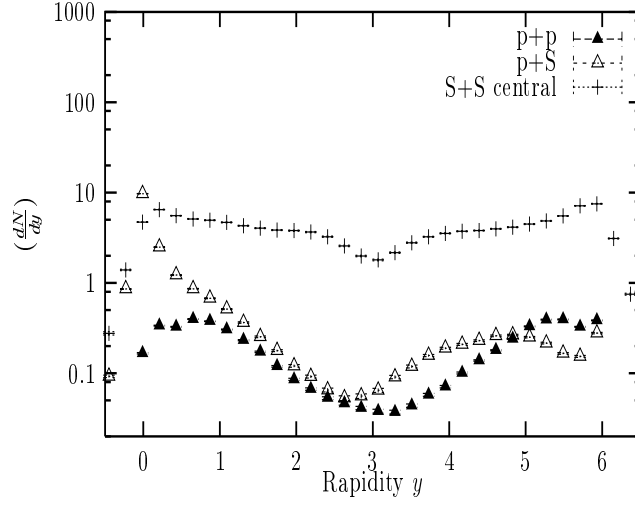,width=8cm,height=7cm}
\end{center}
\vspace*{-3mm}
\caption{Net proton ($p-\bar p$) rapidity distribution in p--p,
p--S and central S--S. 
collisions. Calculated with DPMJET--II.5 without the new diagrams
modifying baryon stopping.
\label{netpnostop1}
}
\end{figure}

\begin{figure}[thb]
\begin{center}
 \psfig{figure=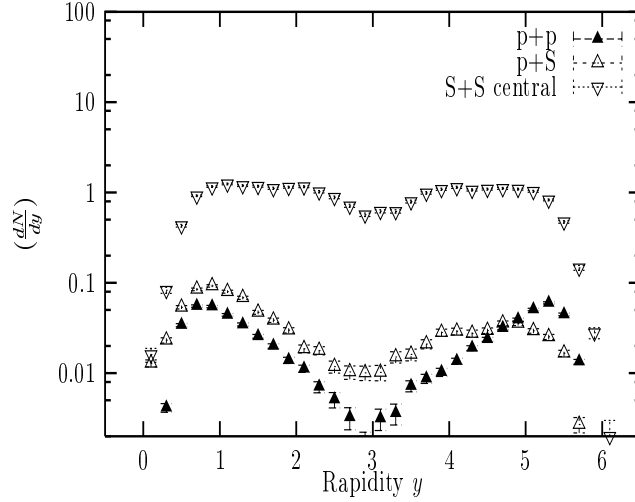,width=8cm,height=7cm}
\end{center}
\vspace*{-3mm}
\caption{
Net $\Lambda$ ($\Lambda-\bar \Lambda$) rapidity distribution in p--p,
p--S and central S--S. 
collisions. Calculated with DPMJET--II.5 without the new  GSQBS and
USQBS diagrams
modifying baryon stopping.
\label{netlamnostop1}
}
\end{figure}

We present first the DPMJET predictions at $p_{lab}$ = 200 GeV/c 
 for net baryon rapidity
distributions in the original model without the new GSQBS and USQBS 
diagrams. In Fig.
\ref{netpnostop1} we present the leading 
net proton ($p-\bar p$) rapidity
distribution $dN_p/dy -dN_{\bar p}/dy$ in p--p,
 p--S and central S--S  collisions.
 In Fig.
\ref{netlamnostop1} we present the 
net  $\Lambda$ ($\Lambda-\bar \Lambda$)  rapidity
distribution $dN_{\Lambda}/dy - dN_{\bar \Lambda}/dy$ in p--p,
 p--S and central S--S  collisions. In p--p collisions,
 which at the given collision energy are hardly modified by the new
 USQBS  diagram, 
we observe a dip at central rapidity. This dip is also
 present in DPMJET without the new GSQBS and USQBS  diagrams 
 in p--S and central A--A collisions. This disagrees to the data
 presented in the next Figures.

\begin{figure}[thb]
\begin{center}
 \psfig{figure=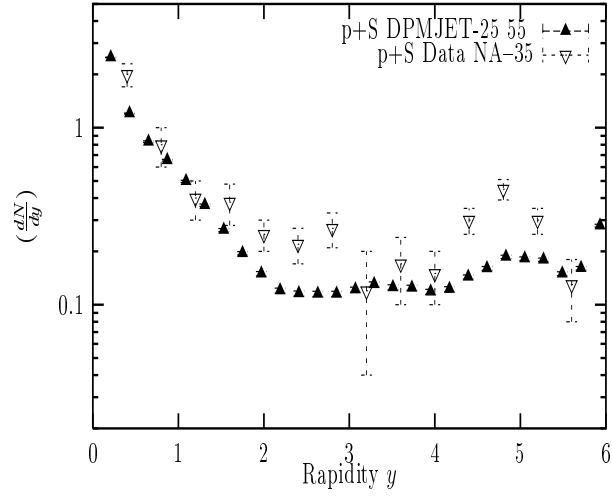,width=8cm,height=7cm}
\end{center}
\vspace*{-3mm}
\caption{Net proton ($p-\bar p$) rapidity distribution in p--S
collisions. 
The DPMJET--II.5 results are compared with data
\protect\cite{Alber98}.
\label{psnetp25}
}
\end{figure}

\begin{figure}[thb]
\begin{center}
 \psfig{figure=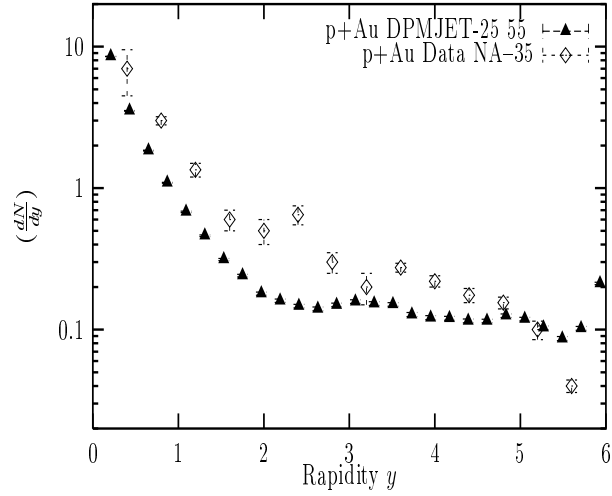,width=8cm,height=7cm}
\end{center}
\vspace*{-3mm}
\caption{
Net proton ($p-\bar p$) rapidity distribution in p--Au
collisions. 
The DPMJET--II.5 results are compared with data
\protect\cite{Alber98}.
\label{paunetp25}
}
\end{figure}

In Fig. \ref{psnetp25} and \ref{paunetp25} we compare the net--proton
distributions according to the full DPMJET--II.5 model with data in p--S
and p--Au collisions \cite{Alber98}. 
Now the dips at central rapidity are filled in the
model, we observe like in the data at central rapidity a flat
net--proton rapidity distribution. 

\begin{figure}[thb]
\begin{center}
 \psfig{figure=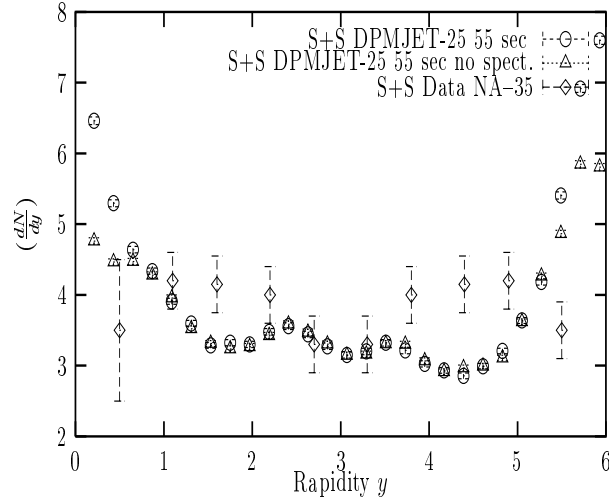,width=8cm,height=7cm}
\end{center}
\vspace*{-3mm}
\caption{Net proton ($p-\bar p$) rapidity distribution in central 
S--S
collisions. 
The DPMJET--II.5 results are compared with data
\protect\cite{Alber98}.
DPMJET--II.5 is used with secondary interactions of comovers and
we present the DPMJET results with and without proton spectators.
\label{ssnetp25}
}
\end{figure}

In Fig. \ref{ssnetp25} we compare the full DPMJET--II.5 model in the 
version  with secondary interactions with data
on net--proton production in central S--S collisions. We give
the DPMJET results with and without spectator protons
(evaporation protons from the residual nuclei). Also here the dip
at central rapidity in the model 
has disappeared (compare to Fig.\ref{netpnostop1}),
however, the agreement
to the data \cite{Alber98} is not perfect. 
There is a significant disagreement in
the fragmentation regions of the two nuclei, especially with the
DPMJET version including the spectator protons. 
The spectator evaporation protons are apparently not included 
in the data.

\begin{figure}[thb]
 \begin{center}
  \psfig{figure=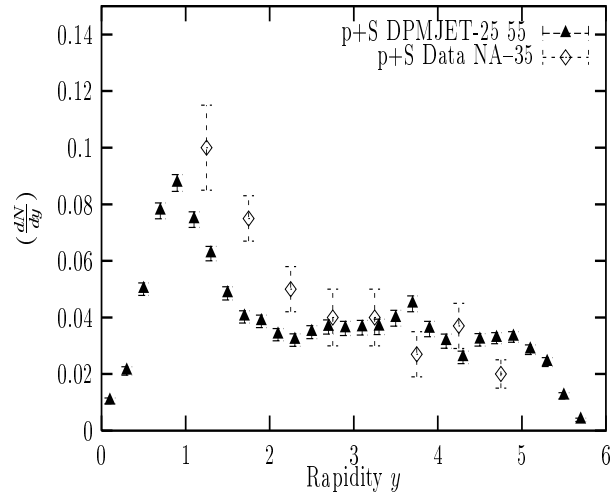,width=8cm,height=7cm}
 \end{center}
 \vspace*{-3mm}
 \caption{Net $\Lambda$ ($\Lambda-\bar \Lambda$) 
 rapidity distribution in p--S
 collisions. 
 The DPMJET--II.5 results are compared with data
 \protect\cite{Alber98}.
 \label{netlamps25}
 }
 \end{figure}

In Fig. \ref{netlamps25}  we compare the full
model to data on net--$\Lambda$ production in p--S 
collisions. In Fig.\ref{netlamps25} the dip at central rapidity has
disappeared in the model. 
%
 
%

\section{Properties of the model in the highest energy region, the
consequences of the baryon stopping mechanisms at Cosmic Ray energies.}

In all plots in this Section we present the DPMJET--II.5 results for the
two versions of the model discussed already in Section II.C: 

(i)With only the new diagram for baryon stopping GSQBS, this version is
characterized in the plots by 50, this version corresponds in its high
energy behaviour approximately to the previous version of 
the model DPMJET--II.4.

(ii)With both new diagrams for baryon stopping GSQBS and USQBS, this
version is characterized in the plots by 55, the hight energy
behaviour of this version is new, at high energies the baryons carry 5
to 10 percent less energy than in version 50, correspondingly the
mesons carry more energy than in version 50.

 We start as example with one plot 
 where the difference between the two versions
 of the model is rather insignificant.



\begin{figure}[thb]
\begin{center}
 \psfig{figure=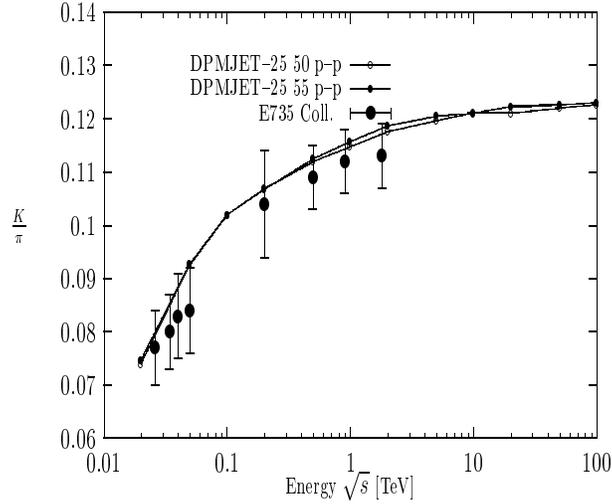,width=8cm,height=7cm}
\end{center}
\vspace*{-3mm}
\caption{K/$\pi$ ratios in $pp$ or  $p\bar p$ collisions as function of
the cms energy $\sqrt s$. The DPMJET--II.5 calculation is
compared with data collected from the E735 Collaboration at
Fermilab \protect\cite{Alexopoulos92}.
\label{dpm222kdpi}
}
\end{figure}

With rising energy the fraction of strange hadrons rises slowly.
In Fig.\ref{dpm222kdpi} the K/$\pi$ ratio according to DPMJET--II.5
for $ pp$ or $p\bar p$ collisions is compared to data from the
E735 Collaboration
\cite{Alexopoulos92}.

We observe similar insignificant differences between both
versions of the model in plots of the average transverse
momentum $<p_{\perp}>_{ch}$ as function of the collision energy
or  of the average charged multiplicity $<n_{ch}>$ as
function of the collision energy.
%

%


Next we discuss plots with significant differences between  the
versions with (55 in the plots) and without (50 in the plots)
the USQBS mechanism.

Following for instance the basic discussion of ~\cite{gaistext},
we introduce a variable $x_{lab}$ similarly to Feynman--$x_F$, but this
time in the lab--frame :

\begin{equation}
x_{lab} = \frac{E_i}{E_0}
\end{equation}

$E_i$ is the lab--energy of a secondary particle $i$ and $E_0$ is the 
lab--energy of the projectile in a h--nucleus collision.
We introduce $x_{lab}$ distributions $F(x_{lab})$ :

\begin{figure}[thb]
\begin{center}
 \psfig{figure=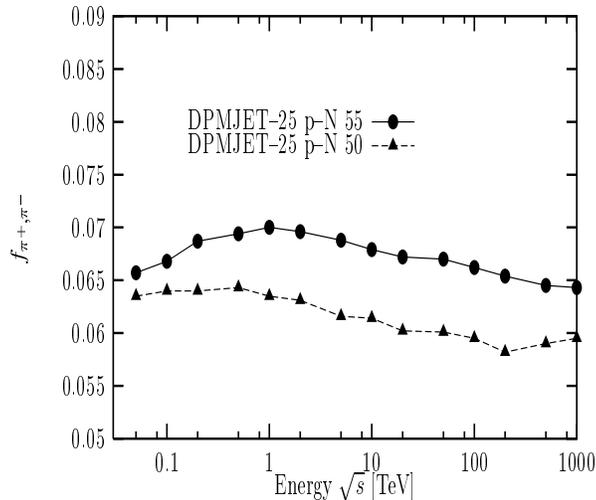,width=8cm,height=7cm}
\end{center}
\vspace*{-3mm}
\caption{Spectrum weighted moments for pion production in  p--N
 collisions as function of the (nucleon--nucleon)
cms energy  $\sqrt s$.
\label{dpm222fpi}
}
\end{figure}

\begin{equation}
F_i(x_{lab}) = x_{lab}\frac{dN_i}{dx_{lab}}
\end{equation}

We note that the Feynman--$x_F$ distribution at positive $x_F$ in the
projectile fragmentation region is a very good approximation to
the $x_{lab}$ distribution. 

The cosmic ray spectrum--weighted moments in p--A collisions 
are now defined as moments of the $F(x_{lab})$ :
\begin{equation}
f^{p-A}_i = \int^{1}_0 (x_{lab})^{\gamma -1}
F^{p-A}_i(x_{lab})dx_{lab}
\end{equation}
Here $-\gamma \simeq$ --1.7 is the power of the integral cosmic
ray energy spectrum and $A$ represents  the target nucleus.
\begin{figure}[thb]
\begin{center}
 \psfig{figure=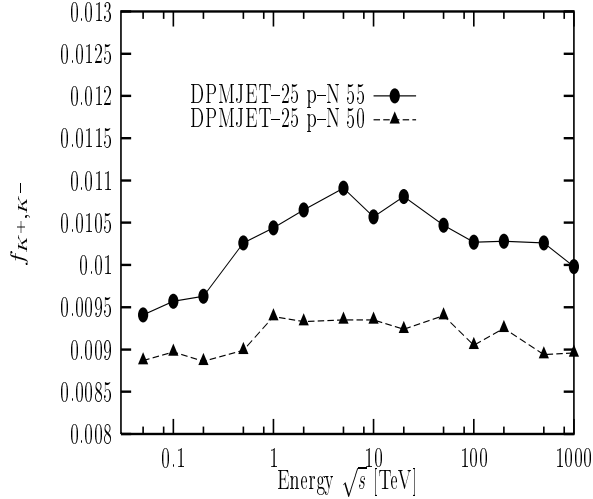,width=8cm,height=7cm}
\end{center}
\vspace*{-3mm}
\caption{Spectrum weighted moments for Kaon production in  p--N
 collisions as function of the (nucleon--nucleon)
cms energy  $\sqrt s$.
\label{dpm222fk}
}
\end{figure}

The spectrum--weighted moments for nucleon--air collisions,
as discussed in ~\cite{gaistext}, determine the
uncorrelated fluxes of energetic particles in the atmosphere.

We also introduce the energy fraction $K^{p-A}_i$ :

\begin{equation}
K^{p-A}_i = \int^{1}_0 
F^{p-A}_i(x_{lab})dx_{lab}
\end{equation}
As for $x_{lab}$, the upper limit for $K$ is 1 in h--nucleus
collisions.

In Figs.\ref{dpm222fpi}  
we present the spectrum weighted moments
summed over pions of both charges 
 in  p--N collisions as function of
the cms energy $\sqrt s$ per nucleon. In Figs.\ref{dpm222fk} 
the
moments are given for charged Kaon production also in 
p--N collisions. 

\begin{figure}[thb]
\begin{center}
 \psfig{figure=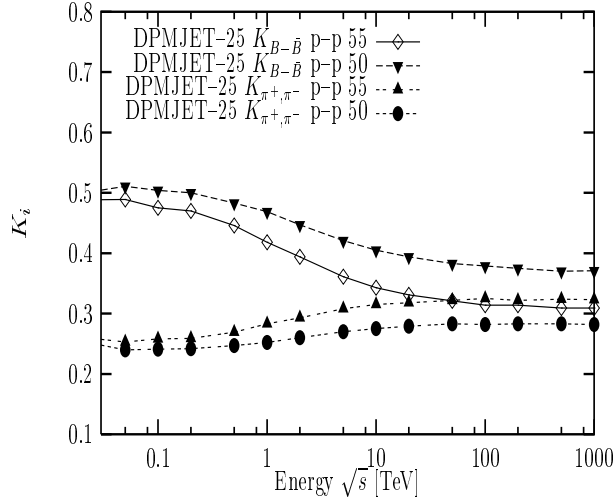,width=8cm,height=7cm}
\end{center}
\caption{Laboratory energy fractions  for net baryons (baryon minus
antibaryon) $B-\bar B$,
baryons  $B$ and
charged pion  production in p--p 
 collisions as function of the (nucleon--nucleon)
cms energy  $\sqrt s$.
\label{dpm25kbppa}
}
\end{figure}

\begin{figure}[thb]
\begin{center}
 \psfig{figure=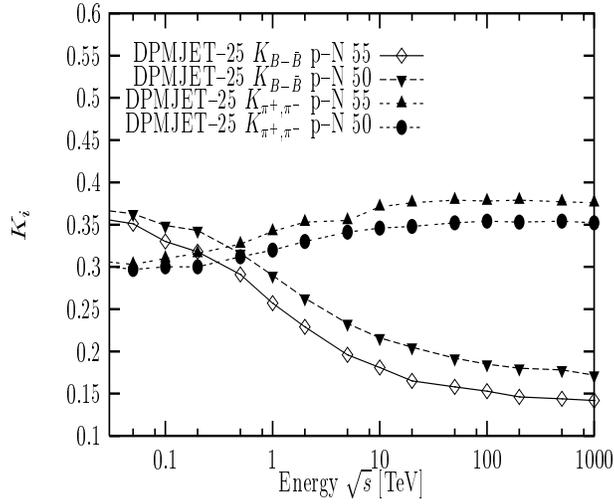,width=8cm,height=7cm}
\end{center}
\caption{Laboratory energy fractions  for net baryons (baryon minus
antibaryon) $B-\bar B$,
baryons  $B$ and
charged pion  production in  p--N
 collisions as function of the (nucleon--nucleon)
cms energy  $\sqrt s$.
\label{dpm25kbpair}
}
\end{figure}

In Fig.\ref{dpm25kbppa} and \ref{dpm25kbpair}  
we present again for $pp$ anf p--N
collisions the energy fractions K  for net baryons $B-\bar B$ (baryon
minus
antibaryon) and charged pion production. The energy fraction
$K_{B- \bar B}$ is always smaller than the energy fraction of
baryons $K_B$. The
difference between both is the energy fraction going into antibaryons 
$K_{\bar B}$  which is equal to the energy
fraction carried by the baryons which are produced in baryon--antibaryon
pairs. 

 \clearpage

We find in DPMJET--II.5  
all average values characterizing hadron production:
the average transverse
momenta, 
the charged multiplicities 
 and the moments in Figs. \ref{dpm222fpi},
\ref{dpm222fk}, \ref{dpm25kbppa} and \ref{dpm25kbpair} 
to change smoothly with
energy in most cases just like the logarithm of the energy.

So far, we have not found any experimental 
data to favour either the versions of DPMJET--II.5
(i or 50) with only GSQBS or (ii or 55) with GSQBS
and USQBS.  
 Clearly, the version prefered by theoretical prejudices is
(ii or 55). In this version we have a better Feynman scaling of meson
distibutions and spectrum weighted moments and we have a faster decrease
 with the collision energy of the energy fractions into secondary
 baryons.


%
%
\section{Summary}

In the present paper we discuss new diquark breaking mechanisms,
which lead to a better agreement with experimental data on
baryon stopping in hadron--nucleus and nucleus--nucleus
collisions in multichain fragmentation models. This is
demonstrated using a particular Monte Carlo version DPMJET--II.5
of the Dual Parton Model. A  new diquark
breaking diagram USQBS is emphasized, which is unimportant at
the energies of the present baryon stopping data. This new
diquark breaking diagram becomes very important at high energies
and it changes significantly the extrapolation of the
multistring models to
Cosmic Ray energies. At the same time the new diquark breaking
mechanisms do not spoil the good agreement of the model to
leading proton Feynman $x$ distributions at present accelerator
energies.

%
\vspace{1cm}
\section*{Acknowledgements}
The author acknowledges   
the fruitful collaborations with P.Aurenche, G.Battistoni, 
F.Bopp, M.Braun, A.~Capella,
R.Engel, A.Ferrari, C.Forti, K.H\"an\ss gen, K.Hahn, 
I.Kawrakov, C.Merino,  N.Mokhov, H.J.M\"ohring, C.Pajares,
D.Pertermann, S.Ritter, S.Roesler, P.Sala  and
J.Tran~Thanh~Van on the Dual Parton Model in general.

%
%
%
\bibliographystyle{physrev}
\bibliography{dpm11}
 
 \clearpage

%
 \end{document}